# Mini-LANNDD:
# A Very Sensitive Neutrino Detector To Measure $sin^2(2\theta_{13})$[*]

*David B. Cline and Stanislaw Otwinowski, University of California Los Angeles, Department of Physics and Astronomy, Box 951447 Los Angeles, California 90095-1547 USA*

*Franco Sergiampietri, INFN-Pisa, Istituto Nazionale di Fisica Nucleare, Via Livornese, 129156010 San Piero a Grado (PI), Italy*

**Abstract**

The ICARUS Detector at the LGNS will carry out a sensitive search for a $sin^2(2\theta_{13})$. We describe a small version for the LANNDD proton decay detector (70kT Liquid Argon) to measure $\nu_\mu \to \nu_e$ in a low energy or off-axis neutrino beam. We find an optimal detector size is 5 kT and at a distance of about 700 km from a high-energy neutrino source. This detector uses the ICARUS method.

1. *Introduction.*
2. *Reach of ICARUS T3000.*
3. *Mini-LANNDD.*
4. *Reach of Mini-LANNDD.*

**Introduction**

The ICARUS detector concept is now well tested and plans for a 3 KT detector at the LGNS are in place by 2005-2006 [1]. This detector with the CNGS neutrino beam will provide the most sensitive search for $\nu_\mu \to \nu_e$ that is currently planned. To go beyond this sensitivity, a new detector will be required in a different (lower energy or off-axis) beam. In this note we discuss a possible 5 kT detector based on the ICARUS method, and a smaller version of the proposed LANNDD detector. LANNDD (Liquid Argon and Neutrino and Nuclear Decay Detector) is a 70 kT magnetized detector for the WIPP Site at Carlsbad, New Mexico. Figure 1 shows a schematic of LANNDD and Figure 2 shows a schematic of the detector located at the WIPP Site[2].

In this note we provide a brief discussion of Mini-LANNDD, a 5kT detector based on the LANNDD concept. We emphasize here the use to search for $\nu_\mu \to \nu_e$ beyond the ICARUS [3] range to the level of $sin^2(2\theta_{13})$ of 0.005.

---

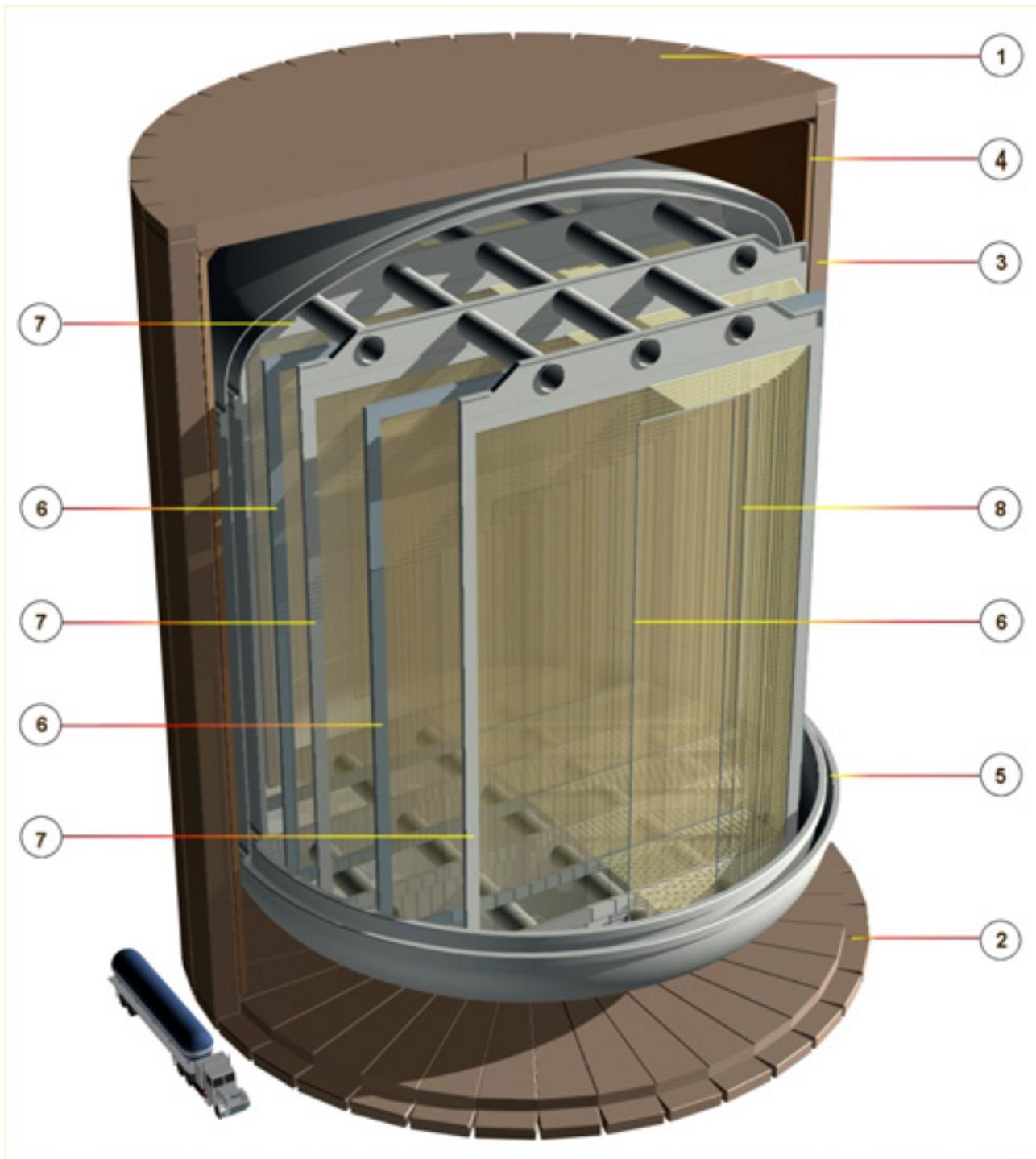

*Figure 1*. LANNDD: 70,000 Tons Magnetised Liquid Argon Time Projection Chamber.
Preliminary sketch:

1) Top iron end cap
2) Bottom iron end cap
3) Iron yoke barrel
4) Solenoid coil
5) Cryostat
6) Cathode planes
7) Wire chambers
8) Drift field electrodes.

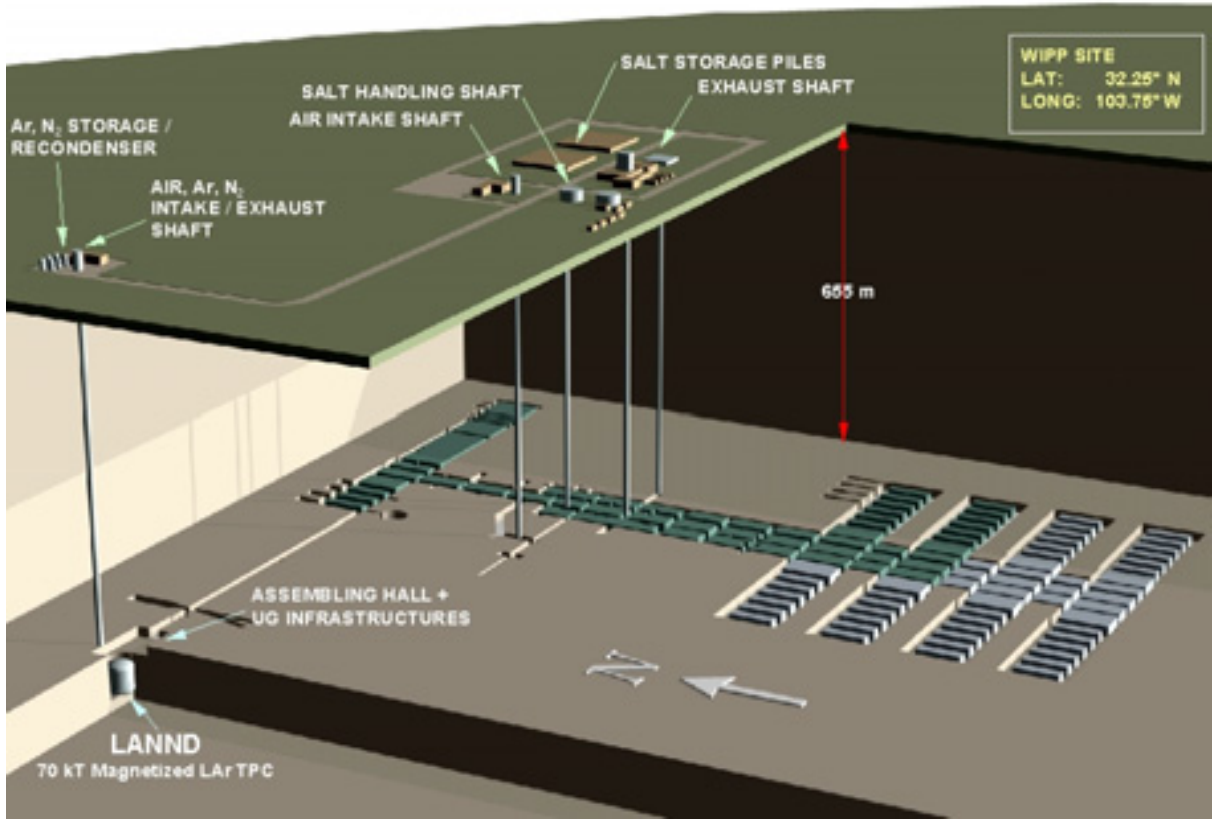

*Figure 2.* The WIPP underground site near Carlsbad, New Mexico.

**2. Reach of ICARUS T3000.**

The ICARUS T600 will be installed soon in the LNGS - most likely in Hall B. A proposal has been submitted to INFN and the USA DOE to construct on additional 2400 tons of ICARUS [4]. In this proposal on analysis of the limit on $sin^2(2\theta_{13})$ that could be reached after 8 years of operating with the CNGS beam – Table 1, lists the events rate for different channels as well as the expected $\nu_e$ background. A detailed analysis by Andre Rubbia [5] and his group indicates that 0.02 could be reached to 90% C.L.. A schematic of the limit that one can reach compared to the background is given Table 2. We see that a value of 0.01 is optimal and with a great deal of effort [or a very good detector] (Table 3 and 4) like mini LANNDD; 0.005 might be reached.

**Table 1.** ICARUS T3000 in CNGS beam. (≈730 km). Rates from $\nu_\mu \to \nu_e$ oscillations in three family mixing (for $\Delta m^2_{23} = 3.5 \times 10^{-3}$ eV$^2$ and $\theta_{23} = 45°$). Rates are normalised to 20 kty. (8 years running on "shared beam").

| $\theta_{13}$ (degrees) | $sin^2(2\theta_{13})$ | $\nu_e$ CC | $\nu_\mu \to \nu_\tau$ $\tau \to e$ | $\nu_\mu \to \nu_e$ | Total | $S/\sqrt{B}$ |
|---|---|---|---|---|---|---|
| 9 | 0.095 | 79 | 74 | 84 | 237 | 6.8σ |
| 8 | 0.076 | 79 | 75 | 67 | 221 | 5.4σ |
| 7 | 0.058 | 79 | 76 | 51 | 206 | 4.1σ |
| 5 | 0.030 | 79 | 77 | 26 | 182 | 2.1σ |
| 3 | 0.011 | 79 | 77 | 10 | 166 | 0.8σ |
| 1.5 | 0.003 | 79 | 77 | 2.5 | 158 | 0.2σ |

**Table 2.** Optimal $sin^2(2 \cdot \theta_{13})$ determination in $\nu_e$ background.

"$\nu_e$ flux issue"

$$\frac{R(\nu_\mu \to \nu_e)}{R(\nu_e)} = \frac{A}{B} sin^2 2\theta_{13}$$

A = function of [$\theta_{23}$, osc. prob. ($4_E$)]

B = Ratio $\frac{(\nu_e)}{(\nu_\mu)}$ in beam

$$\frac{R(\nu_\mu \to \nu_e)}{R(\nu_e)} \cong 1 \to \text{signal} \cong \text{Bkg}$$

Difficult to go far below this level. Must know Bkg very well!

A ~ _ optimal oscillation distance
B ~ _ $10^{-2} \to (0.4 - 0.8) \times 10^{-2}$

NuMi     CNGS??

$sin^2(2\theta_{13}) \sim 10^{-2}$ is optimal: to go below this, the value of Bkg should be well known.

### 3. Mini-LANNDD.

We have made a very preliminary design of a 5 kt LANNDD like detector to use in a NuMi like beam (or off axis CNGS beam). Fig. 3 shows one design including the Cryostat whereas Fig.4 shows a cut through of the detector. We assume 5 m drift length. In Table 5 we give some preliminary parameters of the detector.

### 4. The reach of Mini-LANNDD.

Using some value given to us by Adam Para [6] [7] we show the signal and background for Mini-LANNDD with 20 kty exposure – Table 3 and Table 4. Neutral current will be unimportant. Mini-LANNDD can reach ultimate limit possible in the NuMi beam (or off axis CNGS) of 0.005.

**Table 3.** Mini-LANNDD in NuMi medium energy beam. ($\approx$730 km). Rates from $\nu_\mu \to \nu_e$ oscillations in three family mixing (for $\Delta m^2_{23}$ =3.5x10$^{-3}$ eV$^2$ and $\theta_{23}$ =45°). Rates are normalised to 20 kty. (4 years running on beam of 3.8x10$^{20}$ pot).

| $\theta_{13}$ (degrees) | $sin^2(2\theta_{13})$ | $\nu_e$ CC | $\nu_\mu \to \nu_\tau$ $\tau \to e$ | $\nu_\mu \to \nu_e$ | Total | $S/\sqrt{B}$ |
|---|---|---|---|---|---|---|
| 9 | 0.095 | 60 | - | 382 | 442 | 49$\sigma$ |
| 8 | 0.076 | 60 | - | 304 | 364 | 32$\sigma$ |
| 7 | 0.058 | 60 | - | 234 | 294 | 30$\sigma$ |
| 5 | 0.030 | 60 | - | 121 | 181 | 23$\sigma$ |
| 3 | 0.011 | 60 | - | 44 | 104 | 5.7$\sigma$ |
| 1.5 | 0.003 | 60 | - | 11 | 71 | 1.4$\sigma$ |

**Table 4.** Mini-LANNDD in NuMi mediun energy beam. ($\approx$730 km). Rates from $\nu_\mu \to \nu_e$ oscillations in three family mixing (for $\Delta m^2_{23}$ =3.5x10$^{-3}$ eV$^2$ and $\theta_{23}$ =45°). Rates are normalised to 20 kty. (4 years running on beam of 3.8x10$^{20}$ pot) [6].

| Detector position | $\nu_e$ (no osc) | $\nu_e$ (osc=100%) | $\nu_\mu \to \nu_e$ $sin^2(2\theta_{13}) = 0.005$ | $\nu_e$ bckg | |
|---|---|---|---|---|---|
| 0n axis | 13460 | 4000 | 20 | 60 | $\approx$3$\sigma$ |
| off axis 5km | 5420 | 3000 | 15 | 24 | $\approx$3$\sigma$ |
| off axis 10 km | 1380 | 1300 | 8 | 4.4 | $\approx$4$\sigma$ |

We wish to thank Adam Para and Debbie Harris for requesting this study.

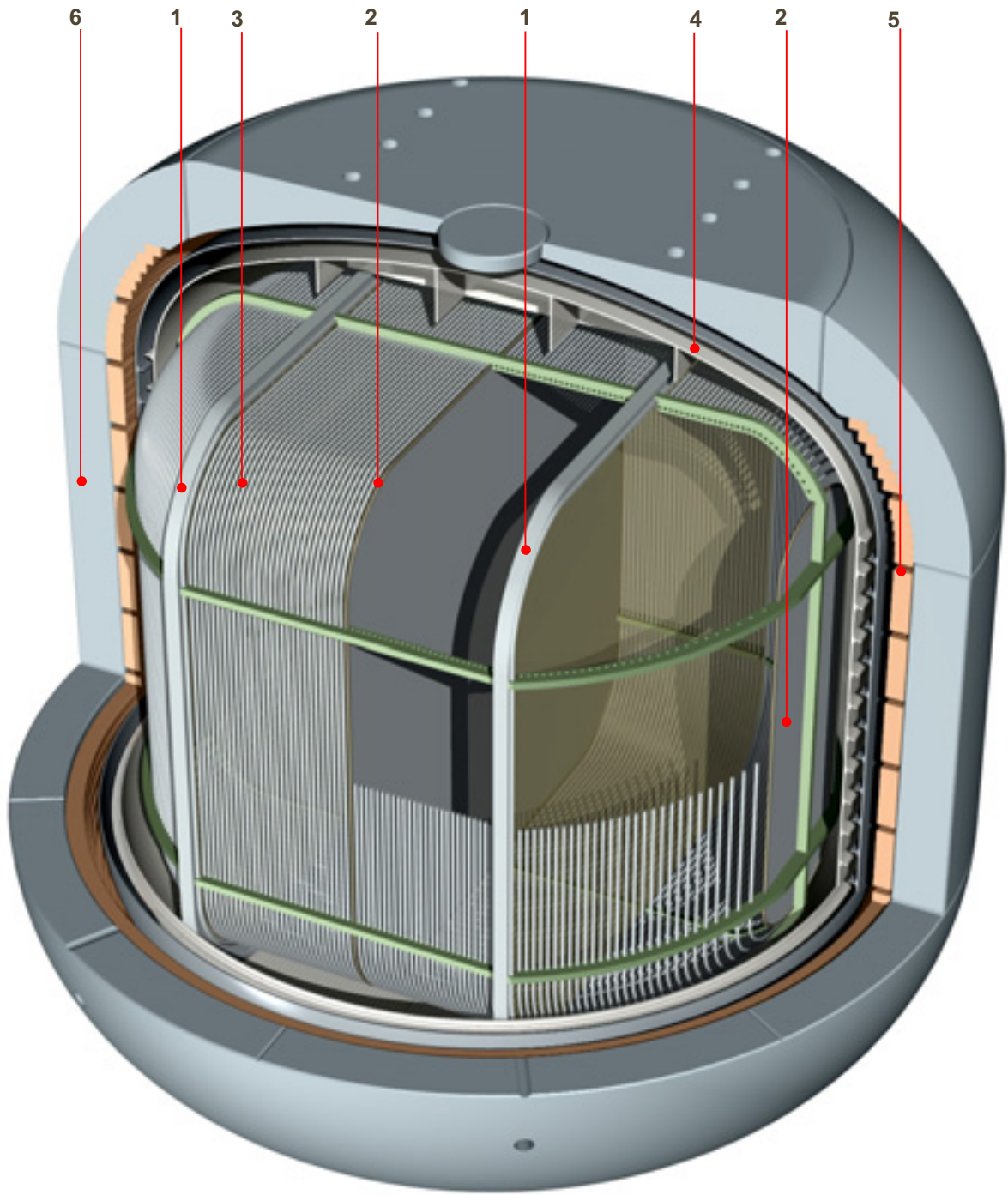

Figure 3. Mini LANNDD: 5,000 Tons Magnetised Liquid Argon Time Projection Chamber.
Preliminary sketch:
1) Wire chambers
2) Cathode planes
3) Drift field electrodes
4) Cryostat
5) Solenoid coil
6) Iron yoke.

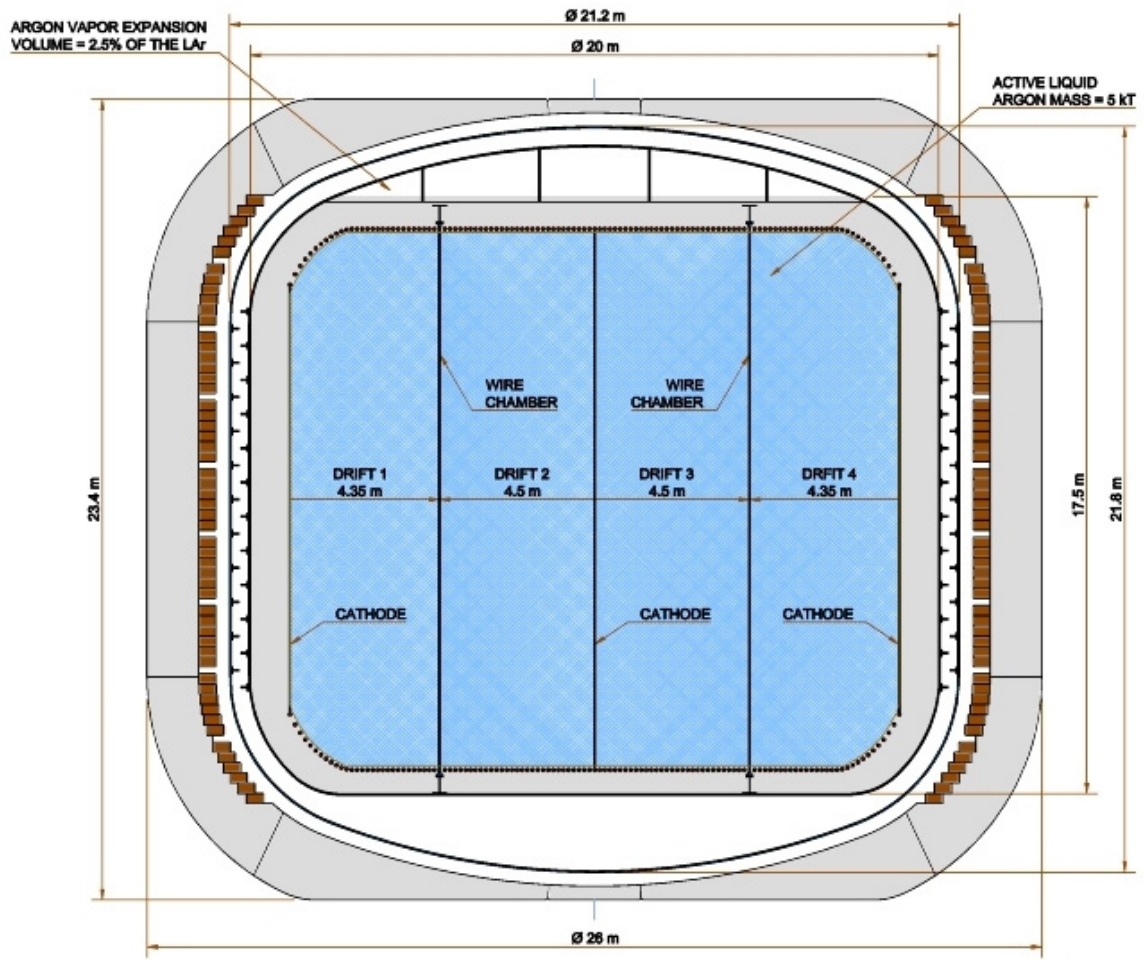

Figure 4. Mini LANNDD: Transverse cross-section.

**Table 5.** Mini–LANNDD - Parameters

| | |
|---|---|
| Active Liquid Argon Volume | 3'587.6 m³ |
| Maximum W×H×D | 17.7 m ×15.6 m ×11.2 m |
| Active Liquid Argon Mass | 4'986.8 Ton |
| Total Liquid Argon Volume | 5'216.8 m³ |
| Total Vapour Argon Volume | 130.6 m³ |
| Vapour-to-Liquid Ratio | 2.5 % |
| Number of Drift Regions | 4 |
| Drift Lengths | 2×4.35 m + 2×4.5 m |
| Maximum Required High Voltage | 225 kV |
| Number of Cathode Planes | 3 |
| Number of HV Feedthrough | 1 |
| Number of Wire Chambers | 2 |
| Number of Read-out Wire Planes | 8 |
| Number of Read-out Wires | 85'160 |
| Maximum Wire Length | 16 m |
| Maximum Wire Capacitance | 320 pF |
| Number of Signal Feedthrough Chimneys | 32 |
| Maximum Signal Cable Length | 5 m |
| Maximum Signal Cable Capacitance | 215 pF |
| Number of Analog-Digital Processing Crate Pairs | 148 |
| Heat Input  a) Radiation | 1.64 kW |
| b) Conduction through cables and supports | 1.3 kW |
| Total | 2 kW |
| LN$_2$ Consumption | 1.03 m³/d |
| Magnetic Field | 0.5 T |
| Total Coil Copper | 5'030 Ton |
| Power Dissipation | 6.2 MW |
| Iron Yoke Mass | 26'440 Ton |